%% file: main.tex
\documentclass[submission,copyright,creativecommons]{eptcs}
\providecommand{\event}{ThEdu'21} % Name of the event you are submitting to
\usepackage{underscore}           % Only needed if you use pdflatex.

\usepackage{graphicx}
\usepackage{mathpartir}
\usepackage{alltt}
\usepackage{amsmath}

\begin{document}

\title{Evolution of SASyLF 2008-2021}
\author{John Tang Boyland%
\thanks{Work done in part while the author was in residence at
  Northeastern University, Boston, MA}
\institute{%
  University of Wisconsin-Milwaukee
  % \streetaddress{P.O. Box 784}
  {Milwaukee}, Wisconsin, USA}
  % \postcode{53201}
\email{boyland@uwm.edu}
  % \orcid{0000-0002-1048-8850}
}

\def\titlerunning{Evolution of SASyLF}
\def\authorrunning{Boyland, John}

\maketitle

\begin{abstract}
\input abstract.tex
\end{abstract}

\input intro.tex
\input changes.tex
\input future.tex
\input related.tex
\input conc.tex

\bibliographystyle{eptcs}
\bibliography{main}

\end{document}

%% file: abstract.tex
SASyLF was released in 2008 and used as a proof assistant in courses
at several universities.  It proved itself useful and has continued to
be used, and each iteration of use has encouraged further development:
fixing bugs and adding enhancements.  This paper describes how SASyLF
was developed while keeping true to its purpose. Most notable are
making substitutions explicit, support of ``and'' and ``or,'' support for mutual
and lexicographic induction, and IDE support.

%% file: intro.tex
\section{Introduction}\label{sec:intro}

SASyLF (Second-order Abstract Syntax Logical Framework) is a proof
assistant based on LF using higher-order abstract syntax
(HOAS)~\cite{pfenning/elliott:88hoas}, but limited to
second-order, for notational simplicity.
SASyLF was introduced in
2008~\cite{aldrich/simmons/shin:08sasylf} by Aldrich, Simmons and
Shin.
The project has a webpage (\url{sasylf.org})
hosted at Carnegie Mellow and a page at \textsf{github}
(\url{https://github.com/boyland/sasylf/}).

SASyLF has been
used in courses based on
Pierce's \textit{Types and Programming
  Languages}~\cite{pierce:02types} at Carnegie Mellon University,
University of California at Los Angeles, University of
Wisconsin-Milwaukee, ETH Zurich, and Northeastern University.
The initial paper discusses some of the responses collected from
students, including that most felt that SASyLF helped them learn how
to write natural langauge proofs.
The repeated interaction with new
users exposes previously unknown defects, underscores known
mis-features and leads to enhancement requests.
This
paper reports on the work completed (as of SASyLF release 1.5.0)
%, made public in December 2020
and the plans for the future.

The stated purpose of SASyLF was to provide a gentler learning curve
to proof mechanization for students within a type theory course,
and secondarily for researchers.  The design philosophy of SASyLF was
(and is)
to follow conventions of paper proofs as closely as possible, but also
to make all the steps explicit, so that errors could be easily
pinpointed and expressed in terms of these high-level concepts.

For example, when describing the structure of the language supporting
the type system, SASyLF follows paper conventions by having the user
specify a context free language, for example for the simply-typed lambda calculus
with a ``unit'' constant:
\begin{quote}
  \begin{minipage}{2.5in}
    \begin{tabular}{ccl}
      $e$ &::=& \(\mathtt{\lambda} x \texttt{:}\tau \cdot e \) \\
      &$|$& $x$ \\
      &$|$& $e$ $e$ \\
      &$|$& \texttt{()}
    \end{tabular}
  \end{minipage}\hfill
\begin{minipage}{2.5in}
\begin{verbatim}
syntax
  e ::= fn x : tau => e[x] 
     | x
     | e e
     | "(" ")"
\end{verbatim}
\end{minipage}
\end{quote}
On the left, we give a typical presentation in a paper, and on the
right, we give the equivalent in SASyLF (from
Figure~1~\cite{aldrich/simmons/shin:08sasylf}).
In this example, we see that expressions (``\texttt{e}'') have four
forms:
lambda expressions, variables, application and the unit constant
(``\texttt{()}'').  The parentheses for the unit constant in SASyLF have to be
quoted since parentheses have special meaning.  The other symbols
(e.g., ``\texttt{:}'') do not need to be quoted, but may be.
The square brackets in SASyLF
(``\verb|[.]|'') used in the definition of lambda expressions express
that the body of the lambda may have uses of variable ``\texttt{x}''
in it.  SASyLF supports higher-order abstract syntax
(HOAS~\cite{pfenning/elliott:88hoas}) so that the user doesn't have to
define complex name-based substitution rules.

Another way that shows how SASyLF follows paper conventions for
specifying type systems is in the inference rules.  For example,
consider the
declaration of the typing judgment and one of its rules:
\begin{quote}
  \begin{minipage}{2in}
    \begin{mathpar}
      \textrm{(typing)} \qquad \Gamma \vdash e : \tau

      \inferrule*[Right={T-App}]
                {\Gamma \;\vdash\; e_1 : \tau' \rightarrow \tau \\\\
                 \Gamma \;\vdash\; e_2 : \tau'}
                {\Gamma \;\vdash\; e_1 \, e_2 : \tau}
    \end{mathpar}
  \end{minipage}
  \hfill
  \begin{minipage}{2.75in}
\begin{verbatim}
judgment has-type: Gamma |- e : tau
assumes Gamma

Gamma |- e1 : tau’ -> tau
Gamma |- e2 : tau’
------------------------- t-app
Gamma |- e1 e2 : tau
\end{verbatim}
\end{minipage}
\end{quote}
The figure on the left is a typical presentation which shows the form
of the typing relation and then uses natural-deduction-style inference
rules.  On the right, we express the same idea in SASyLF
(adapted from 
Figure~2~\cite{aldrich/simmons/shin:08sasylf}).
The only
significant difference is that in SASyLF, the fact that
``\texttt{Gamma}'' includes bindings that may be used in the
expression is declared using ``assumes.''

The Curry-Howard Correspondence
% (also known as the Curry-Howard Isomorphism)
highlights the
connection between computation and logic.  In particular, a proof can
be seen as a computational object, and \emph{vice versa}.
The correspondence is often used to highlight the logical foundations
of programming, especially to mathematicians.
In contrast,
SASyLF encourages the dual viewpoint by using programming as a
metaphor for proof construction.  The purpose is to make the less
familiar world of logical proofs
accessible to competent programmers.
For example, if a proof uses a lemma, it will use
syntax strongly resembling a function call, and a proof
by induction is expressed with (functional) recursion.
An example of a proof in SASyLF in given in the next section in
Figure~\ref{fig:compare}.

As originally designed, SASyLF incorporated a number of aspects to
help it meet its design goals.   In particular:
\begin{itemize}
  \item The syntax of the terms being studied is described with
    context-free grammars.  Syntactic 
    terms and judgments are described syntactically and parsed using
    GLR parsing~\cite{glr}.
  \item Name binding is supported using Higher-Order Abstract Syntax (HOAS)
    within LF~\cite{pfenning/elliott:88hoas}.
  \item Derivations are produced one step at a time (``let normal
    form'').
  \item Incremental proof construction is aided by the ability to
    leave unspecified part of the proof while continuing elsewhere.
\end{itemize}
%These aspects have proved their worth in many iterations of use in
%instruction.

Seeing all
the struggles that compilers students have
getting a grammar accepted as LALR(1) (or LL(k)), it
is refreshing that users in SASyLF simply need to give a context-free grammar.
Ambiguity need only be handled where it occurs, primarily using
parentheses.  So for example, the following term is ambiguous:
\begin{quote}
  \texttt{fn x => x "(" ")"} \quad must be written\\
  \texttt{fn x => (x "(" ")")} \quad or \\
  \texttt{(fn x => x) "(" ")"} \quad.
\end{quote}
The original SASyLF paper mused the addition of
precedence declarations to handle ambiguity, but this was never
added.  For pedagogical purposes, it seems best to require that all
ambiguity be explicitly avoided with parentheses; it makes it clear that
the various choices are different ASTs.

The use of HOAS has been mostly successful.
The alternative, deBruijn notation, is tricky to
learn and even in its clearest exposition (``locally
nameless''~\cite{chargueraud:11nameless}) requires several technical
lemmas for every use of binding.  In HOAS, students mainly need only be aware
of standard static scoping rules, plus be cognizant of ``alpha
renaming,'' the irrelevance of formal parameter names.  The hardest
parts to explain are the restrictions on the ``assumption rule.''

The most serious problem with HOAS is that certain proof approaches are simply
not possible.  
For instance, it's not possible to express
that all types of variables in a
context have a certain property that is not shared by all
types.
%\footnote{Technically: the only restrictions that can be
%placed are syntactic, and must apply to all contexts.}
%Contexts must not be treated as syntax.
We have started the theoretical work on
extending LF with new features to support the necessary changes to the
meta-language~\cite{boyland/zhao:14lf+tuples} but that work has stalled.
% Further discussion on this topic occurs in Section~\ref{sec:future}.

In sharp contrast to most other proof assistants,
SASyLF eschews the use of ``tactics,'' powerful
techniques for generating a proof automatically.
At the risk of gross generalization,
tactics handle all the
easy aspects of proofs but leave the harder parts.  This is the wrong
approach for 
a novice, who lacks the intuition  for the easier parts, let alone the
harder parts.

So it seems that SASyLF as designed has the right approach for
educational purposes; our approach to changes is to ensure that these
goals are kept in view.
In the next section, we discuss the changes already made to SASyLF,
and in the following section look at changes that are under consideration.

% LocalWords:  SASyLF Boyland mis Lambek GLR LF LALR ASTs HOAS
% LocalWords:  deBruijn

%% file: changes.tex
\section{Changes in SASyLF}

The original paper anticipated further developments.
For reference, they are listed here with a brief comment on to what
extent the anticipation was realized.
\begin{enumerate}
  \item ``[We] are developing an integrated development
    environment that will help both novice and expert users write
    boilerplate code. For example, it will allow students to drag a
    rule into a case analysis and have the syntactic structure of
    the case for that rule generated automatically.'' % (page 2)

    \begin{em}
      Indeed, an Eclipse plugin for SASyLF was initiated within a short
      time and this plugin has undergone numerous improvements.
      There is still no support for ``dragging'' a rule into a case
      analysis, but perhaps better, a ``Quick Fix'' has been provided
      which will populate a case analysis with all missing cases.
    \end{em}
    
  \item ``In the future we plan to support namespaces to allow a file
    to rely on declarations from another file without worrying about
    name clashes.''

    \begin{em}
    The module system is still basic, but does permit a proof to be
    used in another proof without name clashes.
    \end{em}
    
  \item ``We have considered adding `by solve' which would
    automatically search for a derivation, but this poses challenges
    in terms of giving away a derivation to students in an
    educational setting.''

    \begin{em}
      Indeed ``by solve'' does introduce the equivalent of a simple
      tactic into SASyLF.  It was added as a project but hasn't been
      maintained.  Fortunately for educational purposes, it can only 
      find simple proofs.
    \end{em}
    
  \item ``However, there are a few checks that are not yet implemented, including the checks for substitution, weakening, exchange, and contraction.''

    \begin{em}
      Substitution, weakening and exchange were all fully checked soon
      after Boyland took over maintenance.
      Contraction remains a reserved word, but nothing can be proved
      ``by contraction'' (the justification is not even accepted by
      the parser). Since contexts in SASyLF are (currently)
      inextricably bound up with variable introduction, and since it
      is not legal to bind the same variable twice (\(\/\Gamma, x:T, x:T
      \vdash R[x]\) is not legal), contraction is not ever applicable.
      The superficially similar situation (\(\Gamma, x:T, y:T \vdash
      R[x,y]\)) can be already be reduced using ``substitution'' to
      (\(\Gamma, x:T \vdash R[x,x]\)).
    \end{em}
    
  \item ``Although some cleanup work is needed, the implementation could eventually serve another educational role: illustrating, through a translation from paper-proof syntax into the LF
    type theory, the formal underpinnings of common notation and some of
    the basic ideas behind LF-based theorem provers.''

    \begin{em}
      We are not aware of any use of SASyLF for this purpose.
    \end{em}
    
  \item ``We believe Part 2B [of POPLmark] will also be feasible in SASyLF; Parts 1A and 2A will require the implementation of mutual induction (not expected to be hard to add since the type theory has already been worked out in Twelf). 

    \begin{em}
      Indeed mutual induction was added early on.  It is \emph{not}
      needed for Part 2A, but is needed for Part 1A with narrowing
      being proved inductively with transitivity of algorithmic
      subtyping.
      Part 2A has been completely mechanized in SASyLF, but Part 1A is
      currently not mechanizable due to the context manipulation
      needed by narrowing.
    \end{em}
    
  \item ``This comment suggests the need for better visualizations
    of proofs—--at a minimum, syntax highlighting, but perhaps output
    in {\LaTeX} or a graphical depiction of a derivation tree. Several
    students felt that a tree-like presentation of a derivation was
    easier to read than the linear format supported by the tool.''

    \begin{em}
      The IDE includes syntax highlighting which is a big help to avoid
      some simple syntax errors, but graphical views have not been
      implemented.  Nor has there been any demand for them, since the
      original paper, to our knowledge.
    \end{em}
    
  \item ``Most \ldots students still did have times when they wished
    they had better feedback, for example real-time feed-back on
    syntax errors and incorrect rule applications, rather than
    continually rerunning the tool, which was `disruptive and
    annoying.' IDE integration could help with such issues.''

    \begin{em}
      Indeed, anecdotally, the IDE has been very helpful to students,
      although some used the command-line version occasionally.
      %Some
      %of the most useful facilities (e.g. filling in missing cases)
      %have support at the command-line as well, to assist those who
      %wish to use other IDEs.
    \end{em}
\end{enumerate}
In the remainder of this section we discuss the changes that actually
have been put into effect, as of SASyLF 1.5.0, starting with Eclipse plugin.

\input plugin.tex

\input comparison.tex

\subsection{Conjunction and Disjunction}

In SASyLF, a theorem or lemma has a single judgment as a result; if
two results are desired they must be packaged together in a single
judgment.  Sometimes a choice result is desired; for example in an
introductory type systems course, students prove ``progress''
with the result that a term is either a value \emph{or} can be
evaluated in a further step.

Conjunction and disjunction can be done easily through the introduction
of new judgments:
\begin{mathpar}
  \inferrule{A \\ B}{A \texttt{ and } B}

  \inferrule{A}{A \texttt{ or } B}

  \inferrule{B}{A \texttt{ or } B}
\end{mathpar}
Ignoring contexts for now, nothing more at the meta-level is needed
for conjunction to work. 
The system as implemented currently creates these judgments
automatically (conjunctions or disjunctions of two or more judgments
at a time) on demand.
%The parsing of conjunctions (and disjunctions) is done
%specially rather than use the general LR parsing, in order to give
%better error messages.
As a special case, the (new) keyword \verb|contradiction| is parsed an
empty disjunction.\footnote{The other use of the new keyword is the
justification ``by contradiction on'' which refers
to an empty case analysis.}

A technical difficulty with disjunctions arises when one
considers that the implicit arguments of a rule includes all free
meta-variables of the premises or conclusion.  In the case of
conjunction, all meta-variables in the conclusion are present in the
premises, but
in the case of a disjunction, meta-variables free in one disjunct but
not in another need to be supplied even when the latter is being used
to satisfy the disjunction. For a simple example, consider a
typical ``progress'' theorem, which proves that a term $t$ is a
value or that is evaluates \( t \to t' \).  The question is what
should $t'$ be for the disjunction if $t$ is a value?

Since meta-variables in SASyLF are never dependently typed, and they
are described
with context-free grammars, the types are always inhabited assuming
that the non-terminal is \emph{productive}, which is a simple recursive check.
Unproductive nonterminals are arguably errors in the specification,
and so SASyLF was updated to mandate that all non-terminals
be productive.
That simple change is enough for the meta-theory to accommodate
conjunctions and disjunctions of
judgments without contexts.

If a conjunction is proved by a lemma or theorem (or inversion), the
syntax requires that the separate judgments be split and named separately,
as seen in in Fig.~\ref{fig:compare} in two cases of inversion.
This obviates the need for an explicit elimination rule for
conjunctions.  Conjunctions are introduced by simply listing the two
parts as in ``\verb|proof by d1, d2|.''

In contrast,
using a disjunction needs case analysis, as seen in this excerpt from the
``progress'' theorem from the
\href{https://github.com/boyland/sasylf/blob/master/examples/iso-recursive-sub.slf}{mechanization of iso-recursive subtyping
from the public repository}:
\begin{quote}
\begin{verbatim}
ns1: t1 value or t1 -> t1' 
    by induction hypothesis on d1
_: t1 t2 -> t' by case analysis on ns1:
    case or e1: t1 -> t1' is
        e: t1 t2 -> t1' t2 by rule E-App1 on e1
    end case
    ...
\end{verbatim}
\end{quote}
The syntax doesn't require that the whole (implicitly created) rule is
given, it is necessary only to provide the particular disjunct being
handled.  Introducing a disjunction is done by proving one of the
disjuncts, or a disjunction of a subset of them.

Joining judgments with contexts interacts with the meta-theory because
the context is not actually part of the judgment.  As a result, a rule
declaration can only produce a judgment with an opaque context
(e.g., \texttt{Gamma}).  The premises must use this context, possibly
adding extra requirements.  For example the typical \textsc{T-Abs}
rule has the form
\begin{mathpar}
  \inferrule*[Right={T-Abs}]{\mathtt{\Gamma, x:T \vdash t[x] : T'}}{\mathtt{\Gamma \vdash
    \lambda x:T . t[x] : T \to T'} }
\end{mathpar}
The meta-level expression of this rule is
\begin{mathpar}
  \inferrule*[Right={T-Abs}]{\Pi{x:\mathtt{t}}.\mathtt{typing}[x,T]
    \to \mathtt{typing}[t[x],T']}
  {\mathtt{typing}[\mathtt{lam}[T,t],\mathtt{arr}[T,T']]}
\end{mathpar}
This form uses
\texttt{typing} for the LF type of the typing judgment,
square brackets \( [ \ldots ] \) for application of arguments to
an LF function or dependent type, and \( \Pi{x:T}.T'[x] \) for dependently-typed
function types (and traditional function type syntax
$T \to T'$ when the result type is independent of the argument type).
The constants \texttt{lam} and \texttt{arr} are the constructors of
lambda expressions and arrow types respectively in the syntax being described.
Here we distinguish the (LF) type of terms \texttt{t} from a
meta-variable and implicit argument to the rule \(t\).\footnote{%
Some readers may notice that, against representation norms, we
don't give $t$ in ``eta-expanded'' form when we pass it to
\texttt{lam}.  Internally, the eta-expanded form is used.}

So the question is how do we represent the result of a hypothetical
canonical forms lemma for arrow types?
\[
\mathtt{t = \lambda x:T . t1[x] \texttt{ and } *, x:T \vdash t1[x] : T'}
\]
In this case, only one judgment uses a context (the equality judgment here
only operates on closed terms), and so we can just pull the entire
context out and generate the conjoined judgment with the single rule:
\[
\inferrule*[Right={\textup{\texttt{Eq+T}}}]
          {\mathtt{t_1 = t_2} \\
            \mathtt{\Gamma \vdash t_3 : T}}
          {\mathtt{t_1 = t_2 \texttt{ and } \Gamma \vdash t_3 : T}}          
\]
Then our example can be rendered as follows where the LF context is
pulled to the outer level:
\[
\Pi{x:\mathtt{t}}.\mathtt{typing}[x,T] \to \texttt{Eq+T}[t,\texttt{lam}[T,t_1],t_1[x],T']
\]
The excess context will not affect the equality judgment in the
premise, since it does
not depend on the context.

A similar approach can be used if multiple judgments use the same
context and the conjunction of the derivations all use the same
derivation, but what if we have a version of the last example where
equality needs the context?
\[
\mathtt{\Gamma \vdash t = \lambda x:T . t1[x] \texttt{ and } \Gamma, x:T \vdash t1[x] : T'}
\]
In this case, pulling out the context to the top would cause the
equality judgment to depend on the bindings.  The solution is to
create a special judgment with a single rule that can add the needed
binding just for typing judgment:
\[
\inferrule*[Right={\textup{\texttt{T*V}}}]
  {\Pi{x:\mathtt{t}}.\mathtt{typing}[x,T_1] \to
    \texttt{typing}[t_2[x],T_3]}
  {\texttt{T*V}[T_1,t_2,T_3]}
\]
Then the desired conjunction can be represented as
\[
\texttt{Eq+T*V}[t,\texttt{lam}[T,t_1], T, t_1, T']
\]
The same approach can be used to conjoin judgments that use different
contexts as long only one them builds on unknown bindings in the LF
context.  This restriction is moot since SASyLF already bans a theorem
(or lemma) or judgment from having two contexts.

Once the meta-theory was handled, what remained was to check the
contextual parts that are not handled by the meta-theory (naming of
contexts) and to provide implicit coercions to hide the existence of
the constructed judgments and rules.

\subsection{Syntactic Sugar}

It is common in presentations of type systems to give multiple letters for
use as meta-variables, for example for types:
\[
S, T \:{::=}\: \texttt{A} \mid S \to T \mid S + T
\]
The grammar for SASyLF had ``space'' for this extension, so it was
added.  Around the same time, short-cut definitions were added so that
one can write
\[
\textrm{id} \:{:=}\: \mathtt{\lambda x. x}
\]
The operator \verb|:=| is used because we are not defining a new
non-terminal, but rather defining a short-hand.

In the same vein, \emph{derived syntax} can be defined.  It differs
from the latter in having non-terminal parameters:
\[
\mathtt{(t_1\texttt{; } t_2) \:{:=}\:
  (\lambda x:\texttt{Unit}. t_2) t_1}
\]
The numbering ensures that the terms are substituted in the correct
order.  The left-hand side must be parenthesized to avoid parsing
ambiguity, and is defensible as one is defining the syntax as a whole.

Derived syntax was easy to add to the system, as it simply added
some new productions to the concrete syntax used in the generalized LR
parser.
%\textsl{[Probably can delete the remainder of this section:]}
%Interestingly, adding this extension revealed that the underlying LF
%implementation wasn't able to handle higher-order functions: lambda
%abstractions with functional type could not be applied.  LF itself supports
%higher-order functions (but not polymorphism), so updating the
%implementation to handle higher-order was straightforward.
%
%An example of the need for higher order can come from defining ``let''
%as derived syntax:
%\[
%\mathtt{(\texttt{let } x \texttt{ = } t_1 \texttt{ in } t_2[x])
%  \:{:=}\: (\lambda x . t_2[x]) t_1}
%\]
%The function that converts from one syntax to the other accepts both
%\(\mathtt{t_1}\) and \(\mathtt{t_2}\), where the second has a function
%type.

\subsection{Induction}

The fundamental danger with recursion is non-termination, and the
corresponding problem with proofs using induction is circular
reasoning.  And so induction can only be permitted if the recursion
can be shown to always terminate.  The standard way to prove
termination is with a ``measure'' that is \emph{reduced} in every call.
The measure needs to be a \emph{well-founded} partial order, that is, it
has no infinite decreasing chains.

In computer science, \emph{structural} induction is the fundamental
induction technique in proofs.  It uses the structure of terms: an
inductive call with one term is a reduction of the current call if the
new term is a sub-term of the current one.  Since all terms are (by
definition) finite, this measure is guaranteed to be well-founded.
SASyLF uses structural induction with both syntax terms and judgments.
A premise used in the rule used to prove judgment instance is a
sub-term.

This basic concept is complicated by higher-order terms, from binders
and from assumptions.  We generalized SASyLF's rules of structure
induction on HOAS to follow those of Twelf.  The Twelf User
Manual~\cite{pfenning/schuermann:02twelf}
describes restrictions sufficient to ensure soundness.

This discussion of structural induction permits us now to
discuss the user-visible extensions to SASyLF.
One of the earliest additions to SASyLF was mutual induction.
Theorems (and lemmas) can be connected into a mutual induction group
through the keyword ``\verb|and|'' and then (forward) references within
the group are allowed as long as inductive argument is reduced.
And backward references are allowed additionally even if the inductive
argument is
unchanged.

An immediate issue is that how does one theorem ``know'' what the
induction argument of another theorem is?  In the original SASyLF,
induction was only indicated in a form of case analysis.  But the proof
of a mutually inductive theorem may not need any case analysis, and
forcing a case analysis would needlessly complicate the proof.
And so mutual inductive theorems were assigned their first argument as
the inductive argument, by default, if there is no explicit induction
declaration.

%The original SASyLF uses ``by induction hypothesis'' to handle uses of
%induction. This had to be generalized for mutual induction; the choice
%was to continue to support the special syntax but only for immediate
%(self) induction, and to use the name of the theorem otherwise in the
%same way that theorems are used normally.  This syntax underlines the close
%connection between induction and recursion.

Some time after mutual induction was added, new syntax was added to
permit induction to be declared separate from case analysis:
\begin{quote}
  \texttt{use induction on }$d$
\end{quote}
For backward compatibility (for the vanishing small SASyLF community),
the implicit induction on the first argument was preserved.
%Inadvertently in a later code cleanup, the implicit declaration of
%induction was
%generalized to apply to all theorems, even those with explicit uses of
%the ``induction hypothesis.''
This implicit induction has been deprecated in
SASyLF 1.5.0; all mutually inductive theorems and all theorems using
induction (explicit or implicit) must have an explicit
declaration of induction to avoid a warning.

The new syntax for declaring induction gave ``space'' to extend
structural to lexicographical induction (as seen in the
\href{https://github.com/boyland/sasylf/blob/master/examples/cut-elimination.slf}{example proof
of ``cut-elimination''}):
\begin{quote}
  \texttt{use induction on A, d1, d2}
\end{quote}
This powerful induction technique permits induction to seek a
reduction in \texttt{d1} if \texttt{A} stays the same, and further to
look for a reduction in \texttt{d2} if the first two are unchanged.
What makes this reduction measure powerful is that if \texttt{A} is
reduced, then \texttt{d1} and \texttt{d2} are unrestricted; they can
get arbitrarily bigger.

Another common induction situation is where arguments are swapped in a
use of induction.  Proofs of transitivity of algorithmic subtyping
often have such a situation when handling contravariant type
constructors such as in function types.  SASyLF was extended to
support ``unordered'' induction (as seen in the \href{https://github.com/boyland/sasylf/blob/master/examples/LF.slf}{example proof of
composition of hierarchical substitution in LF}):
\begin{quote}\def\{{\char123}\def\}{\char125}\def\a{\(\mathtt{\alpha}\)}
  \texttt{use induction on \{\a0, \a2\}, s2}
\end{quote}
This example uses lexicographical induction in which the first
argument is an unordered induction of the two ``simple types''
(\(\mathtt{\alpha 0}, \mathtt{\alpha 2}\)), and the second is a
substitution judgment (\texttt{s2}).
Unordered induction requires that under some permutation of the
elements, every element reduces or stays the same, with at least one
element reducing.
The possibility for permutation makes it strictly more powerful than
the similar induction schema in Twelf, which is just a special case of
lexicographic induction.

The LF hereditary substitution composition example involves mutual
induction as well.  All theorems in
a mutual induction group must declare induction of the same form (same
number, groupings and type).

\subsection{Special cases}

A useful programming idiom (especially in imperative programming) is to
handle a special case before a more involved computation, because then
the more involved computation does not need to take into the
special case that has already been handled.  A similar situation can
happen with proofs: there may be a case for one of the inputs to a
theorem that if handled ahead of time, the rest of the proof can be
easier.  These special cases are unremarkable in informal proofs, but
don't fit well into a formulation based on functional programming.

\emph{Partial case analysis} in SASyLF 1.5.0 can be used to
handle such situations.
\begin{quote}
\begin{verbatim}
do case analysis on ...:
  ...
end case analysis
\end{verbatim}        
\end{quote}
A partial case analysis on a subject has the same kind of
cases and the same proof obligation as the surrounding context, but it
need only handle a subset of the possible cases.  Importantly, the
system ``remembers'' that the cases are already handled and so later
in the same context if there is case analysis on the same subject, the
cases that are already handled can (and indeed must) be omitted.
In particular ``by inversion'' (only one case) and ``by
contradiction'' (no cases) may be possible when they would not be,
previous to the partial case analysis.

Partial case analysis was first introduced as a way to reduce case
explosion, but it has served as a useful tool for working around
incompleteness in higher-order unification.  The mechanization of recent
work on iso-recursive
subtyping~\cite{zhou/oliveira/zhao:20iso-recursive} gives an example.
The canonical forms lemma for arrow type says that if we have a value
of arrow type then the value has the form of a lambda abstraction.
The problem is that the typing judgment includes an ``unfolding''
possibility:
\[
\inferrule*[Right={T-Unfold}]
{\mathtt{T = (\mu X . T_1[X])} \\
 \mathtt{\Gamma \vdash t : T}}
{\mathtt{\Gamma \vdash \texttt{unfold}\,(T)\,t : T_1[T]}}
\]
Recall that the SASyLF term \( \mathtt{T_1[T]} \) is represented with LF term application.  The problem
is that there is no MGU for the unification problem \( T_1[T]
\stackrel{?}{=} \texttt{arr}[T_2,T_3] \).  Thus if a SASyLF proof
attempts to do a case analysis on a proof of \(\mathtt{\Gamma \vdash t
  : T_2 \to T_3}\), the possibility of the \textsc{T-Unfold} rule
leads to an ``incomplete unification'' error.

The work-around, is to ``disguise'' the arrow type in an equality
judgment, do a partial case analysis handling the problematic rules
(such as \textsc{T-Unfold}) and then afterwards invert the equality
judgment and proceed as normal.
%\begin{quote}
%\begin{program}
%lemma canonical-forms-ArrowH:
%    forall d: $\mathtt{ * \vdash t : T}$
%    forall e: $\mathtt{T = T_1 \to T_2}$
%    forall v: t value
%    exists $\mathtt{t = \lambda x:T' . t_2[x]}$
%    ...
%end lemma
%\end{program}
%\end{quote}
%First a partial case analysis is done on the typing judgment (with the
%output type safely disguised as \verb|T|) in which we can handle the
%\textsc{T-Unfold} case easily (since the value judgment cannot be
%true) and incidentally also the \textsc{T-Sub} possibility.
%After that partial case analysis, we proceed to invert the type
%equality, ``revealing'' the fact that \verb|T| is actually an arrow
%type and proceed with the rest of the lemma's proof, in which we can
%ignore the \textsc{T-Unfold} and \textsc{T-Sub} cases.  As is often
%the case, the first case analysis after the partial case analysis is
%\emph{not} on the same judgment.
%Rather the proof proceeds by case analysis 
%on the ``value'' judgment.  The ``value'' judgment
%has cases for each form of value; if it's a lambda abstraction, we have
%the needed equality, otherwise, we use the typing derivation (now
%shorn of problematic cases) to show a contradiction.

\subsection{Relaxation}

The notation that SASyLF supports for using judgments from the LF context
only handles the case where the assumption is the last one added, for example:
\[
\inferrule*[Right={T-Var}]{ }
{\mathtt{\Gamma, x:T \vdash x : T}}
\]
Then ``weakening'' can be used to add more assumptions after the one
used.  This semantics is supported by the meta-theory when creating a
judgment instance.

But pattern matching does not proceed as cleanly.  In Twelf, pattern
matching cannot reach into the LF context.  Instead if one needs to
prove a meta-theorem for which a case can be in the LF context, one
needs to place an instance of the theorem being proved in the context
along with the assumption.  This idiom is peculiar to Twelf and not
close to how one does the proof on paper.

Instead SASyLF permits case analysis to operate on the context.
The problem is that it uses the rule which positions the assumption at
the end of the context.  For example, if we are writing a substitution
lemma that does case analysis on \(\mathtt{\Gamma,x:T_2 \vdash t[x] :
  T_1}\), then as well as the case that \texttt{t[x]} is \texttt{x} (which is
handled without needing to reach into \(\mathtt{\Gamma}\)), we have the additional case:
\[
\inferrule*[Right={T-Var}]{ }
{\mathtt{\Gamma', x' : T_1, x:T_2 \vdash x' : T_1}}
\]
The case apparently implies that $\mathtt{\Gamma = \Gamma', x':T_1}$, in other
words, that the new assumption we found was at the end of the LF
context.  In the 
case of the simply-typed lambda calculus, this implication is fine
since assumptions can be freely permuted.
But in some type systems, the assumptions cannot be permuted, for example in
the presence of type variables, as in $F_{\leq}$.

Indeed, one can construct a ``proof'' of a contradiction in SASyLF if
the apparent implication is permitted to have validity
(see \href{https://github.com/boyland/sasylf/blob/master/regression/bad38.slf}{\texttt{bad38.slf} in
the public SASyLF repository}).  Thus case analysis in the context
needs proper support in the meta-theory.  This support is provided in
the concept of ``relaxation.''

The simple idea at the kernel of relaxation is that rather than saying
that the current context is \emph{equal} to the new (smaller) context with
the assumption at the end, the current context \emph{includes} the new
context with the assumption:
\[
\mathtt{\Gamma \geq \Gamma', x':T_1}
\]
Then any judgment defined in the context on the right can be moved
back to $\Gamma$ through \emph{relaxation}, a form of weakening,
although new assumptions are not explicitly added.
% (other than \(\mathtt{x':T_1}\))
In the process of relaxation, the variable $\mathtt{x'}$ is disguised by
being replaced with $\mathtt{t[\ldots]}$ (the parameter \verb|x| is
known to be unused and so can be replaced with anything in scope). 
Furthermore relaxation gives no status to any intermediate assumptions
(the ``\( \mathtt{x:T_2} \)'' in the example).
In general there could many assumptions between the one found in the
pattern analysis and the known assumptions at the end.

The semantics we give here is a stop-gap until contexts are properly
brought into the meta-theory (see the ``Future Work'' section).

\subsection{Making Substitutions Explicit}

As mentioned briefly earlier in this work, and as described in much
more detail in a separate paper~\cite{ariotti/boyland:17where},
``where'' clauses make explicit the implicit substitution of
meta-variables.  This change is the biggest syntactic change to the
original SASyLF but also completely in line with the purposes of the
system: making proofs explicit, especially for pedagogical purposes.
SASyLF does not require that these ``where'' clauses are
given, unless a certain option is turned on. (In the IDE, the option is
on by default.)

As originally implemented, and as described in the paper cited above,
the extension had no effect on the operation of the rest of SASyLF; it
didn't need to be part of the ``trusted core.''  The substitutions
found were strictly \emph{descriptive}.

This aspect is mainly preserved, but in some situations, especially
with inversion, the result of the substitution does not have a
user-provided presentation, \emph{except} in the ``where'' clauses.
For example (from \verb|good40.slf| in the public repository):
\begin{quote}
\begin{verbatim}
theorem gt-anti-refl-use-w2:
    forall n1
    forall d: n1 > n1
    exists contradiction .
    use induction on n1
    use inversion of rule gt-more on d
      where n1 := s n
    p: n > n by theorem succ-cancels-gt on d
    proof by induction hypothesis on n, p
end theorem
\end{verbatim}
\end{quote}
This proof of the anti-reflexivity of \verb|gt| is an inductive proof,
in that it makes a recursive call.  Unusually, it doesn't use an
explicit case analysis.  Instead it uses inversion on the only rule
matching the input.  In the process, we need a name for the natural
number preceding \verb|n1|.  The ``where'' clause gives the name
``\verb|n|'' for this new variable.  This means that the theorem input
\verb|d| has type \verb|s n > s n|.  This judgment is passed to the
theorem \verb|succ-cancels-gt| which can strip the \verb|s| from both
sides of the inequality.  Then the result can be used in the inductive
call (for which \verb|n| is a sub-term of \verb|n1|).

\subsection{Generalizing Inversion}

The example just shown for ``where'' clauses also demonstrates one of
the generalization of inversion added to SASyLF:
inversion can be performed solely for the ``side-effect'' of the
unification that it performs.    The new syntax ``use inversion''
is commonly used on equality judgments.  Previously, one needed to
perform a case analysis which not only was wordy, but it also
increased the depth of the proof and its visual complexity. Now with
``use inversion,'' one can simply invert the equality judgment in place and then
assume the equality in the remainder of the proof.  This approach is much
closer to how one would reason informally.

Internally, the ``use'' syntax is implemented as proving an empty
conjunction, a tautology.  This prevents a proof (case) from ending with a
``use'' line.  The same technique is used for the ``use induction''
syntax explained earlier.

As can also be seen, inversions can have associated ``where'' clauses
(``further work'' in the ``where'' clause
paper~\cite{ariotti/boyland:17where}).   An interaction with another
extension was the ability to invert ``or'' judgments (when only
one possibility remains, usually because a partial case analysis
eliminated the other possibilities).  The ``obvious'' syntax for such
inversions is the clunky form:
\begin{quote}
  \ldots \texttt{ by inversion of or on } $d$
\end{quote}
The sequence of three two-letter keywords, each starting with ``o'' is
visually confusing.  More on this problem below.

Separately, there was the need to invert syntax: if a non-terminal has
only a single (remaining) possibility, inversion is more attractive
than case analysis.  While it is not common to define non-terminals
with only a single possibility, there are good reasons to do so, for
example in our \href{https://github.com/boyland/sasylf/blob/master/examples/LF.slf}{mechanization of LF}, we define a constructor identifier
as a natural number: ``\verb|c ::= n|.''  More commonly, with partial
case analysis, all but one possibility could have already been
handled.  Again there's a desire for inversion rather than requiring a
case analysis.

An inversion on syntax doesn't use rules, and also doesn't yield any
judgments, and so the syntax is simply:
\begin{quote}
  \texttt{use inversion on} \textit{nt} \texttt{where} \textit{nt}
  \texttt{:=} \textit{form}
\end{quote}
Once this syntax was added, it was a short step further to permit it
also for inversion of judgments, making the identification of the rule
optional.  In particular, inversion of an ``or'' judgment can simply by done
``\texttt{by inversion on}'' the judgment instance, which avoids the
unpleasing syntax, which is available in SASyLF 1.5.0 but probably
won't be used too much.

\subsection{Modules}

Requiring all the elements of a proof to be in a single file is
limiting.  Originally, SASyLF even required there to be a single
syntax section, although this restriction was lifted.  Nonetheless,
it's inconvenient to have to (say) include a definition of natural
numbers (and especially all operations and theorems about them) in
every proof that uses them.  It would also be desirable to have a
library of theorems about groups in general that could be applied to
any operation satisfying the group axioms.  Thus it has been a long-term
goal to add some form of modules.

A syntax has been defined in which a SASyLF file can include a module
header, including a name, a sequences of requirements (syntax,
judgments and lemmas) and sequence of provided declarations.  The
requirements are often ``abstract,'' that is, not defined.  When a
module is used, the requirements are substituted with new elements.
However, the syntax has run ahead of the meta-theory: SASyLF 1.5.0
does not support using modules with requirements. But it does support
the use of modules without requirements and a nascent library has
been defined with modules for natural numbers and Boolean values.

An important design goal is that a proof should be syntactically
independent of the modules it uses: all syntax and judgments defined
for the proof need to be defined locally.  So when syntax from a
module is used, it must be locally defined as well.  This gives an
option for an alternate syntax to be used, for example:
\begin{quote}
  \texttt{syntax n = Natural.n ::= 0 | 1+n}
\end{quote}
A similar situation applies to judgments:
\begin{quote}
  \texttt{judgment nat-ne = Natural.notequal:\ n != n}
\end{quote}
Rules and theorems (and lemmas) can be used with name qualification.
In case analysis on an imported judgment, the rule names in the cases
do not need 
to be qualified (and doing so will occasion a warning).
%This behavior
%follows Java's case analysis on enumeration constants.

SASyLF 1.5.0 uses a package system (whose syntax was defined but
ignored in the
original SASyLF) and the \verb|Natural| module
is actually in the \verb|org.sasylf.util| package, and so the uses
described above are not legal unless one includes a local renaming of
the module:
\begin{quote}
  \texttt{module Natural = org.sasylf.util.Natural}
\end{quote}
An advantage of defining a shorthand is that the IDE will use it
with content assistance: If one types ``\texttt{by theorem Natural.ne-}'' and
presses the content-assist key (e.g., control-space), the system will
suggest three possible completions with their theorem headers in a
helper popup window.

% LocalWords:  SASyLF namespaces Boyland LF POPLmark Twelf LaTeX IDEs
% LocalWords:  tex disjunct Eq polymorphism contravariant iso MGU slf
% LocalWords:  succ nt nat ne

%% file: plugin.tex
\begin{figure*}[tp]
  \noindent
  \includegraphics[scale=0.375,trim=25 50 0 25]{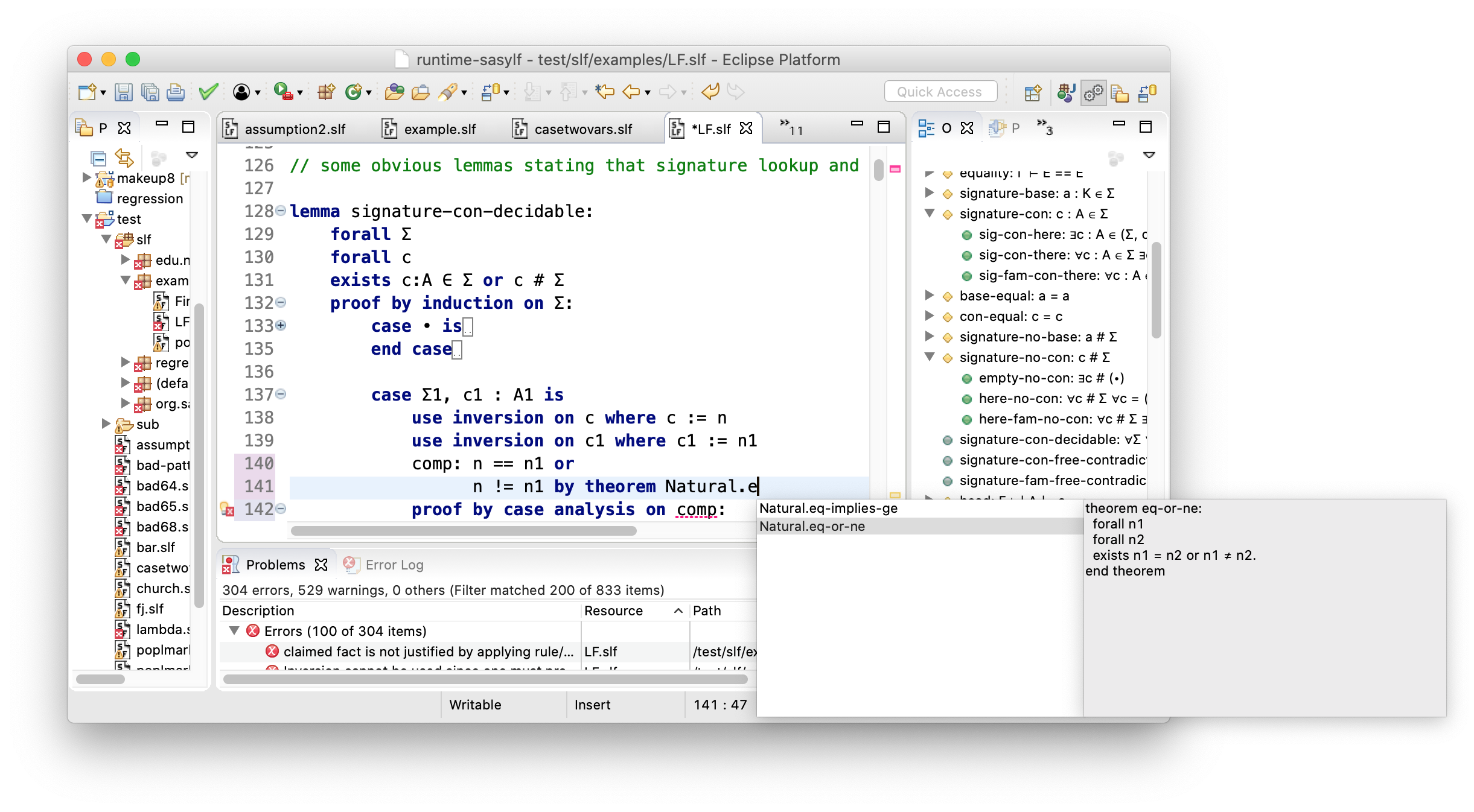}
  \caption{Screenshot of SASyLF 1.5.0 under Eclipse.}\label{fig:plugin}
\end{figure*}
  
\subsection{Eclipse Plugin}

Programmers expect IDE support while coding, and to continue the
identification of proving with coding, SASyLF developers soon provided
support in an Eclipse plugin.  Incrementally, new features have been
added.  While editing, users will notice
\begin{itemize}
\item an outline view,
\item syntax highlighting,
\item parenthesis matching,
\item automated indentation,
\item error and warning markers,
\item folding and unfolding of proof segments,
\item content assist (e.g. lemma name completion).
\end{itemize}
Some of these features can be seen in Figure~\ref{fig:plugin}.
The plugin provides a number of capabilities:
\begin{itemize}
\item Checking proofs;
\item Addition/removal of comments;
\item Quick fixes of many kinds of errors;
\item Finding the definition of a theorem;
\item Changing syntax highlighting colors (``dark'' mode available);
%\item Creating/Deleting proofs, proof packages and proof projects;
%\item Renaming and moving proofs.
\end{itemize}
Unlike common plugins, such as for Java, the syntax is not checked on every
keystroke.  The checking process is still non-incremental and so
is done only on a save, or on demand (e.g., pressing the green
check-mark in the tool bar).

One of the most consequential aspects of the plugin is the ``Quick
Fix'' system. For many errors, the error marker in the left ruler
contains a ``light bulb'' icon (see Figure~\ref{fig:plugin}).
Clicking on that, or using the ``Quick Fix'' keyboard short-cut brings
up a dialog in which one can select a fix.  For missing cases (as this
error is), the plugin will add the cases which the proof is
currently missing.  Indeed this feature is so powerful,
that it is recommended to \emph{not} inform students of its existence
for the first few weeks, so students will write their own case analyses. 

The proof engine of SASyLF (which can be run from the command line) is
architecturally separated from the Eclipse plugin architecture.  
Indeed the ``smarts'' of ``Quick Fix'' 
are available from the command-line, enabling enterprising
students to use SASyLF with \textsf{vim} or IntelliJ.

%% file: comparison.tex
\subsection{Syntactic Comparison}

SASyLF 1.5.0 accepts almost all legal proofs written in the original
syntax, but provides features not originally available.
Figure~\ref{fig:compare}
contrasts Figure 4 from the 2008 paper with a proof more
idiomatic in current SASyLF.
\begin{figure*}\def\K#1{{\bfseries{#1}}}
\begin{minipage}{3.25in}\scriptsize
\begin{alltt}
\K{theorem} preservation: \K{forall} dt: * |- e : tau 
                      \K{forall} ds: e -> e'
                      \K{exists} * |- e' : tau.
dt': * |- e':tau                   \K{by induction on} ds :
  \K{case rule}
    d1 : e1 -> e1' 
    -------------------- c-app-l 
    d2 : e1 e2 -> e1' e2
  \K{is}
    dt' : * |- e' : tau        \K{by case analysis on} dt :
      \K{case rule}
        d3 : * |- e1 : tau' -> tau
        d4 : * |- e2 : tau' 
        -------------------------- t-app 
        d5 : * |- (e1 e2) : tau
      \K{is}
        d6 : * |- e1' : tau' -> tau 
                     \K{by induction hypothesis on} d3, d1
        dt': * |- e1' e2 : tau \K{by rule} t-app \K{on} d6, d4 
      \K{end case}
    \K{end case analysis}
  \K{end case}

  \K{case rule}... // case for rule c-app-r is similar

  \K{case rule}
    d1 : e2 value 
    ---------------------------------------- r-app 
    d2 : (fn x : tau' => e1[x]) e2 -> e1[e2]
  \K{is}
    dt' : * |- e' : tau       \K{by case analysis on} dt :
      \K{case rule}
        d4 : * |- fn x : tau' => e1[x] : tau'' -> tau
        d5 : * |- e2 : tau'' 
        ----------------------------------------- t-app
        d6 : * |- (fn x : tau' => e1[x]) e2 : tau
      \K{is}
        dt' : * |- e' : tau    \K{by case analysis on} d4 : 
          case rule
            d7: *, x:tau' |- e1[x] : tau 
            -------------------------------------- t-fn 
            d8: * |- fn x:tau' => e1[x] : tau' -> tau
          \K{is}
            d9: * |- e1[e2] : tau \K{by substitution on} d7, d5
          \K{end case}
        \K{end case analysis}
      \K{end case}
    \K{end case analysis}
  \K{end case}
\K{end induction}
\K{end theorem}
\end{alltt}
\end{minipage}~%
\begin{minipage}{2.8in}%
  \def\vd{\(\vdash\)}\def\l{\(\mathtt{\lambda}\)}\def\ar{$\to$}%
  \def\w#1{{\normalfont\textsl{#1}}}%
  \def\qw#1{``{#1}''}%
  \scriptsize%
\begin{alltt}
\K{theorem} preservation :
    \w{// Unicode identifiers and operators supported}
    \K{forall} d: * \vd\ t : T
    \K{forall} e: t \ar\ t'
    \K{exists} * \vd\ t' : T \w{// \qw{.} not required}

    \w{// Induction can be declared separately}
    \K{use induction on} d

    \w{// \qw{proof} serves instead of repeating goal}
    \K{proof by case analysis on} e:
        \K{case rule}
            e1: t1 \ar\ t1'
            --------------------- E-App1
            _: (t1 t2) \ar\ (t1' t2)
            \K{where} t := t1 t2 \K{and} t' := t1' t2
            \w{// \qw{where} makes substitution explicit}
        \K{is}
            \w{// \qw{and} allows proving things together}
            d1: * \vd\ t1: T' \ar\ T \K{and}
            d2: * \vd\ t2: T' \K{by inversion on} d
            \w{// \qw{inversion} does a single case analysis}

            d1': * \vd\ t1': T' \ar\ T
              \K{by induction hypothesis on} d1, e1
            \K{proof by rule} T-App \K{on} d1', d2
        \K{end case}

        \K{case rule} ... // case E-App2 is similar

        \K{case rule}
            v2: t2 value
            -------------------------- E-AppAbs
            _: (\l x:T'. t11[x]) t2 \ar\ t11[t2]
            \K{where} t := (\l x:T'. t11[x]) t2
             \K{and} t' := t11[t2]
            \w{// \qw{.} can be used in syntax, e.g., in lambdas}
        \K{is}
            d1: * \vd\ \l x:T'.t11[x] : T''{\ar}T \K{and}
            d2: * \vd\ t2 : T'' \K{by inversion on} d
            d11: *, x:T' \vd t11[x] : T
                 \K{by inversion on} d1
                 \K{where} T'' := T'
            \w{// inversions can cause substitutions too}

            \K{proof by substitution on} d11, d2
        \K{end case}
    \K{end case analysis}
\K{end theorem}
\end{alltt}
\end{minipage}
\caption{Comparing Fig.~4 from~\cite{aldrich/simmons/shin:08sasylf}
  (left) to more idiomatic
  expression in SASyLF 1.5.0 (right).}\label{fig:compare}
\end{figure*}
%In Fig.~\ref{fig:compare}, we show an example of a proof from the
%original paper and compare it with a proof using the new capabilities
%of SASyLF 1.5.0.
The new proof has italicized comments highlighting
the main differences.  Another difference is that the names of
nonterminals and rules is closer to that used in Pierce's
TAPL~\cite{pierce:02types}.

The change that makes the new proof \emph{longer} (otherwise, it is
noticeably shorter) is the addition of optional ``where''
clauses~\cite{ariotti/boyland:17where}.  This addition has an
important explanatory value: it makes explicit how meta-variables
(such as \verb|T''|) are changed by the pattern match.  We found
that students new to proof mechanization often are unaware of how
pattern matching changes the identity of meta-variables.  And not only
novices.
For instance, the answer to exercise 9.3.9 on pages 506 and 507 the
first edition of TAPL
incorrectly presumed that (the equivalent of) \verb|T'| and \verb|T''|
were identical in the \textsc{E-AppAbs} case, when that identification
only happens later when the typing relation for the lambda expression
is inverted.

The other changes here can be seen as conveniences, generalizations and
extensions of the original SASyLF features.

%% file: future.tex
\section{Future Work}\label{sec:future}

In this section, we describe some of ideas for further extension.
We cannot commit to any of these plans, but all have been discussed to
some extent, and seem compatible with the original goals of SASyLF.

\subsection{Support for Equality and Reduction in Theorem Output}\label{sec:future-equality}

Currently, if a theorem is able to show that two terms are the same,
the only way to communicate that fact to clients is to use an equality
judgment.  The clients will then immediately invert the equality
judgment.  This extra step has no counterpart in a textual proof, even
one which carefully lists each step.  Rather the definition of
equality is assumed.  For a while, it seemed that the best way forward
would be to pre-define equality for user-defined types and add extra
support to handle the introduction and elimination.  The handling of
contexts complicates the situation, and also the fact that the
judgment may ``trespass'' on syntax needed for other judgments (e.g.,
an equality judgment used precisely to hide equality from the system
temporarily to avoid the incompleteness of higher-order unification).

The syntax of ``where'' clauses gives a way for equality to be
separated from judgments and given it own space.  It seems attractive
to permit ``where'' clauses to declared in the obligation of a
theorem, as in the following hypothetical example of a uniqueness of
addition theorem:
\begin{quote}
\begin{verbatim}
theorem plus-unique:
    forall d3: n1 + n2 = n3
    forall d4: n1 + n2 = n4
    exists () where n3 := n4
    ...
end theorem
\end{verbatim}
\end{quote}
Then one could write:
\begin{quote}
\begin{verbatim}
use theorem plus-unique on p, p' where n := n'
\end{verbatim}
\end{quote}
Or perhaps even
\begin{quote}
  \texttt{n := n' by theorem plus-unique on p, p'}
\end{quote}
The latter syntax would then also be available to inversion as well.

The ability of a theorem to produce equalities would also make it
easier for a theorem to produce other non-judgment results as well,
such as ``reductions,'' which allow structural induction.
For instance
\begin{quote}
\begin{verbatim}
theorem gt-reduces:
    forall d: n1 > n2
    exists () where n1 > n2
    ...
end theorem
\end{verbatim}
\end{quote}
This theorem would prove that \verb|n2| was a sub-term of \verb|n1|.

%The best workaround to the lack of reduction theorems currently known
%is to pass the judgment to the theorem and use partial case analysis
%with induction to handle the case that the difference is recursive,
%and to use inversion on the judgment once these cases are removed.
%(See for example the decidability of substitution
%(\texttt{subst-m-decidable}) in the mechanization of LF in the public
%repository.)

Adding non-judgment obligations to theorems would affect the rest of
the justification system.  It must be possible, then, for case
analysis (both full and partial) to produce these non-judgment outputs
as well.  The ``\verb|proof|'' syntax would need to include these
extra obligations as well.  And at the end of a sequence of
derivations, we must not only check the last derivation, but also
check the current substitution to see if the equality (or reduction) is
known.  One must also consider whether one can use disjunction with
reduction ot equality.

%Another potential complication would be whether to permit ``or''
%joining of non-judgments.  Consider the following hypothetical
%example:
%\begin{quote}
%\begin{verbatim}
%theorem max-implies-one-eq:
%    forall d: n1 max n2 = n3
%    exists () where n3 := n1 or n3 := n2
%    ...
%end theorem
%\end{verbatim}
%\end{quote}
%It seems best to disallow such disjunction because of the difficulty of
%handling it with anything other than a judgment.  But the syntactic
%merging of ``where'' clauses with derivations would encourage such attempts.

\subsection{Wildcards in Premises}

The simple ability to give ``by unproved'' as the justification for
any step may seem trivial at first, but it is a big help for people
writing proofs.  Rather than leaving an error in place, the form of
the judgment is given.  It can be used and checked at later points.
Proof ``by unproved'' permits proofs to be constructed from the end,
rather than forwards.  First one determines what needs to be proved.
Then one determines what rule (or theorem) would produce the desired
result and what premises it would need.  These can be defined and
justified ``by unproved'' and then the rule application can be checked.

Two students in a type systems class using SASyLF with different
levels of experience with provers 
independently suggested to extend this ability to apply to partial
rule applications.  Currently, if anything is wrong with a rule
application, the whole judgment is flagged as an error, and since
unification is used to check the premises and conclusion together, it
is hard to pinpoint an error to a particular spot.  The proposal is to
permit a premise to be replaced with \verb|_| and then if the
application is otherwise error-free, only the missing part would be
flagged:
\begin{quote}
\begin{verbatim}
da: Gamma |- t1 t2 : T' by rule T-App on _, d2
\end{verbatim}
\end{quote}
The system may even have a suggestion as to the type of the
missing premise(s).
%This syntax could replace the need for the
%``unproved'' keyword:
%\begin{quote}
%\begin{verbatim}
%da: Gamma |- t1 t2 : T' by _
%\end{verbatim}
%\end{quote}
This idea seems within the capabilities of the existing engine,
where each kind of justification would need to decide whether to
support such holes.  Presumably case analysis and weakening would not
need to support holes, but rule/theorem application and conjunction
creation would.%
\footnote{%
SASyLF 1.5.1 has implemented this requested feature through the ``Quick
Fix'' system.}

One of the students also suggested permitting such wildcards within a
term in case one is not quite sure what is proved, as in the following
example:
\begin{quote}
\begin{verbatim}
da: Gamma |- t1 t2 : _ by rule T-App on d1, d2
\end{verbatim}
\end{quote}
Again the idea is that an error (or warning) would be generated with
information about the missing information.  At an extreme level, one
could write:
\begin{quote}
\begin{verbatim}
da: _ by rule T-App on d1, d2
\end{verbatim}
\end{quote}
Here the user is not showing any indication of what the result should
be at all.  From the first public release, SASyLF already in some
cases will infer the actual result and print it in an error message, depending
on the error encountered.  So this ability seems well within the
spirit of the tool.

\subsection{Generalization of Contexts}

The basic idea for handling contexts has not changed from the
beginning: a bound variable is associated with a single syntactic production
of the context, and this production is associated with an
``assumption'' rule of a judgment that assumes the context.

The fact that an identifier is known to be a bound variable is due to
it occurring inside square brackets in the grammar (only variables may
thus appear), and the nonterminal in which the variable name occurs by
itself in a production indicates the syntactic restriction on the
variable.

The production of the context nonterminal for a variable has a single
occurrence of a variable, a single (recursive) instance of the context
nonterminal, optionally some other nonterminal occurrences, and any
number of ``noise'' terminals.

The context nonterminal must have a single production with only
terminals, and otherwise can only have variable productions, one for
each syntactic variable form.

The
assumption rule, unlike all other rules anywhere in the system
includes a single production of the context nonterminal in the conclusion.
This production must be the production for a variable, and must
comprise the most unrestricted form of that production.  The
assumption rule is
permitted no premises.  The rest of the rule can use no nonterminals
not occurring in the context production, and must have at least one
use of each nonterminal occurring as well as exactly one use of the
variable.

For instance, in the case of the simply-typed lambda-calculus, one can
define:
\begin{quote}
\def\w#1{{\normalfont\textsl{#1}}}%
\begin{alltt}
syntax
  t ::= lambda x:T . t[x] \w{// \texttt{x} is a variable}
     | t t
     | x      \w{// \texttt{x} is of nonterminal \texttt{t}}

  G ::= *     \w{// terminals only}
     | G, x:T \w{// variable production for \texttt{x}}
      
judgment typing: G |- t : T
assumes G

  ------------- T-Var
  G, x:T |- x:T  \w{// verbatim use of variable production} 

  \(\vdots\)
\end{alltt}
\end{quote}
This information is used in the conversion of forms into LF.
Each explicit binder is converted into two levels of abstraction in LF.
So for example the form
\begin{quote}
\begin{verbatim}
G, x1: T -> T' |- ...
\end{verbatim}
\end{quote}
is converted to
\[
\Pi{x_1:\texttt{t}}.\Pi{d_1:\texttt{typing}[x_1,\texttt{arr}[T,T']]}.
% \Pi{x_2:\texttt{t}}.\Pi{d_2:\texttt{typing}[x_2,\texttt{Top}]}. \ldots
\ldots
\]
The second abstraction's formal (here $d_1$) is never used in the
internal form, and so one can use LF's arrow type short-hand:
\[
\Pi{x_1:\texttt{t}}. \, \texttt{typing}[x_1,\texttt{arr}[T,T']] \to \ldots
\]
From this conversion, it can be seen why we need an assumption rule for
the variable production: it is needed for representation of the
context.  Similarly, the fact that the assumption rule cannot mention
other nonterminals (not in the variable production) is clear for the
same reason of representation, they would be unbound.  Premises for
the rule would be irrelevant since the assumption is a hypothetical,
not a proof.  The fact that the context nonterminal needs a
terminals-only production reflects the fact that the empty context has
no information.  Multiple terminals-only productions would be
confusing since they would not be distinguished in the meta-theory.
But the remaining restrictions in
the context and assumption rules are potential points for extension.
% future-context.tex
Lifting these restrictions would somewhat increase the expressiveness
of SASyLF without requiring any substantial extensions in the
meta-theory.  We now turn to a far-reaching extension.

\subsection{Contexts as Syntax}

One of the restrictions that was discussed above is that a nonterminal
must either be identified as a context or as syntax.  In the former
case, it must not be an object of ``assumes'' and it can serve as a
binder of variables.  In the latter case, any judgment that mentions it
must have a declaration that it ``assumes'' the context nonterminal,
and aside from highly restricted ``assumption'' rules, all uses in
conclusions of rules must be unrestricted.

This restriction can be lifted aside from two difficulties, one that
could be handled withing the current meta-theory but the other causing
a major change.  The first, minor difficulty is how to handle variable
productions, e.g., ``\verb|G, x:T|,'' because if the structure is not
a context, then the variable is not bound, and cannot be used.  The
easy, but unsatisfying answer us that the variable can be simply ignored;
it carries no information. For example suppose we define a judgment on
types that don't use named variables: \verb|T ground|.  Then with the
restriction lifted we could defined a judgment that a context uses
only ground types:
\begin{quote}
\begin{verbatim}
judgment contextground: G ground

  -------- G-Empty
  * ground

  G ground
  T ground
  --------------- G-Var
  (G, x:T) ground
\end{verbatim}
\end{quote}
%Thus far, no meta-theoretical changes are needed.

The main difficulty concerns when a single nonterminal is
treated as a context and as syntax in the same context.  Some
situations cannot be permitted: since a judgment that ``assumes'' a
context does not actually receive information about the context, it
cannot use a judgment that treats it as syntax:
\begin{quote}
\begin{verbatim}
G ground  //! cannot be permitted
G, x:T |- t[x] : T'
----------------------------- T-AbsWrong
G |- lam x:T . t[] : T -> T'
\end{verbatim}
\end{quote}
Indeed such combinations if permitted would render weakening and
exchange unsound in general.
The other way around is a situation in which a judgment that does
\emph{not} assume the context has a premise which does.
This situation contains some of the aspects discussed below, but is
less interesting because the context cannot be used by any of the
other syntactic members of the judgment.

The truly difficult mixing comes in a theorem which has premises which
assume the context and those that do not, for instance:
\begin{quote}
\begin{verbatim}
theorem ground-typing-implies-something:
    forall d: G |- t : T
    forall g: G ground
    exists ...
end theorem
\end{verbatim}
\end{quote}
Since the second premise treats \verb|G| syntactically, the theorem
cannot assume that it is represented in the LF context of the
theorem.  Instead, \verb|G| must be an implicit parameter of the
theorem.  The other premise is represented using the context as an
abstraction, or rather as a series of abstractions depending on the
size of \verb|G|.  LF does not have a way to represent a
variable-length series of abstractions.  It cannot be represented by a single
abstraction with a compound argument because LF provides no way to
extract pieces from compound values.
Preliminary work to extend LF with
variable arity~\cite{boyland/zhao:14lf+tuples} was never completed.
%That work, or something similar would be needed as the
%meta-theoretical basis for permitting using contexts also as syntax.
%Assuming the work was complete, yielding a consistent meta-theory for
%so-called ``@LF'', extending SASyLF would be possible.
% future-context2.tex

\subsection{Parameterized Modules}

With SASyLF 1.5.0, we finally have support for modules, albeit without
parameters.  The syntax and basic idea of module parameters have already
been laid out.  The module can ``require'' parameters, and then
another proof can instantiate the module with parameters

More details need working out before it makes sense to embark on
designing a new set of library modules using parameters: finding the
best way to express something, how can the system be made practical,
making sure that the result has a sound meta-theory.
For all these reasons, development of modules has been slow.

\subsection{Subtyping in Syntax}

In Pierce's TAPL, values are routinely given as a sub-grammar of
terms.  SASyLF does not support this idiom; if the
examples from TAPL are typed in, they are syntactically legal, but
generate ambiguous parses wherever something could be a value or a
term.
Updating SASyLF to support sub-grammars was identified as a desirable
extension in the first partial solution to the original POPLmark
challenge~\cite{aydemir/et.al:05poplmark}.

In this case, as opposed
to some other desirable extensions, the meta-theory of ``refinement
types'' for LF is already completed~\cite{lovas/pfenning:07refinement}.
As part of the formalization, it uses intersection types (or rather
``intersection sorts'').  Indeed, due to pattern matching, a
meta-variable may found to be an element of two sub-grammars (sorts) at once.
This is rather awkward in SASyLF because the name of a meta-variable
is intrinsically bound up in its type: what do we do with (say)
\(\mathtt{t'_2}\) when it is found to be a value?  We would need to update
the system to remember what sorts a nonterminal is (currently) known
to inhabit.  The current sorts would affect which productions are
needed for case analysis, perhaps generalizing the information used in
partial case
analysis.  The implementation of the meta-theory would also require a
major update to support sorts.

\subsection{Co-induction}

Proofs in SASyLF assume that syntax and judgments are finite; that
structures are defined through a least fixed-point of grammar
production and rule application.  Generalizing this assumption is
useful for the support of coinduction.  Kozen and
Silva~\cite{kozen/silva:17coinduction} give a good introduction to how
coinduction should be supported in informal proofs.  This article
should be a good guide for extending SASyLF with coinduction,
translating the category theory
(e.g., the article defined
a coinductive datatype as an element of the final coalgebra of a
polynomial endofunctor on \textbf{Set}) into practical rules.

The biggest obstacle to supporting coinduction over coinductive
datatypes is the need to support circular/infinite structures.  SASyLF
would need syntax to create them, and the internal engine would need a
major rewrite to support these structures, in particular defining
unification on coinductive structures with variables.  Furthermore,
since these structures are typed in LF, it is necessary to
extend LF to operate on ``co-terms.''  It is not clear whether LF is
sound on co-terms.
%\textsl{Has someone done it?  Apparently Coq's type
%  system would be undecidable with ``positive'' coinductive data
%  types.}

% LocalWords:  SASyLF pre subst LF iso mis equi algorithmically andL
% LocalWords:  arity ctx TAPL POPLmark coinduction Kozen coinductive
% LocalWords:  coalgebra endofunctor Coq's

%% file: related.tex
\section{Related Work}

In this section, we review some of the other education proof systems
particularly interested in helping students construct their own
proofs.

Lurch~\cite{carter/monks:13lurch} is
one example of a system with a similar mission as SASyLF, except in
the field of mathematics.  It compares most similarly with SASyLF in
aiming to check proofs written in the idioms of paper proofs.  It
works as a ``word processor'' similar to using SASyLF in an IDE.  It
provides the standard tools of a rich text editor and indeed recently
adds the ability to edit formulas in a simplified text-based
equation system (or even in \LaTeX).  The particular ability
of Lurch is that when parts of 
the text are selected as ``meaningful,'' Lurch will check them.  So if
a mathematical statement such as ``\(\forall x . x^2 \geq 0\)'' is
``meaningful'', it will expect this statement to be followed by a
proof, which is checked.  It uses colored icons to indicate which
statements are proved, definitely wrong or simply unproved.

%The RISC ProgramExplorer~\cite{schreiner:11explorer} which is an
%outgrowth of the ProofNavigator~\cite{schreiner:08navigator} 

AXolotl~\cite{cerna:19axolotl} is a puzzle game for students to
learn how to generate proofs using quantified rules.
The researcher (Cerna) recognized that students often struggle with
unification and substitution in rule application and includes several
aids to help students bridge the conceptual gap.  Recently the tool
has been ported to work as a mobile
application~\cite{cerna/et.al:20mobile}.

The Students' Proof Assistant~\cite{schlichtkrull:19SPA} takes a
different approach.  Rather than a stand-alone tool, it is embedded
within Isabella/HOL. Unlike the standard Isabelle proof assistant, it
keeps the entire tree of the a proof visible to the student working on
a proof.  This feature makes it comparable to SASyLF, despite the fact
that it still supports tactics (being within the Isabelle system).
The researchers agree that it's important for students learning to
write proofs to see structure of the proof being created.

People experienced in writing proofs can find writing the simple parts
of proofs tedious while people just introduced to proving need to work
through these details because everything is new.  Paradoxically, the
expert needs more assistance than the newcomer (albeit only for
productivity); in fact the assistance can be a hindrance to the
student learning how to prove.

%% file: conc.tex
\section{Conclusion}

In conclusion, SASyLF 1.5.0 carries on the vision of the original
SASyLF paper while firming up the foundation, and providing additional
convenience.  Some of that convenience (e.g. ``quick fixes'' in the
IDE) are best not highlighted at the beginning of a course introducing
SASyLF, so as to ensure that
important learning steps are not skipped.  We anticipate that SASyLF
will continue to develop as long as it remains useful pedagogically.